\documentstyle[prl,aps,twocolumn,amsfonts]{revtex}
\def\identity{{\Bbb I}}                      
\def\R{{\Bbb R}}                             
\def\Z{{\Bbb Z}}                             
\def\C{{\Bbb C}}                             
\newcommand{\gmu}{\gamma_{\mu}}             
\newcommand{\gnu}{\gamma_{\nu}}             
\newcommand{\dmunu}{\delta_{\mu,\nu}}       
\newcommand{\gone}{\gamma_{1}}              
\newcommand{\gtwo}{\gamma_{2}}              
\newcommand{\gfive}{\gamma_{5}}             
\newcommand{\cbar}{\overline c}             
\newcommand{\A}{A}                          
\newcommand{\B}{B}                          
\newcommand{\OA}{\overline{A}}              
\newcommand{\OB}{\overline{B}}              
\newcommand{\OG}{\overline{G}}              
\newcommand{\oa}{\overline{a}}              
\newcommand{\ob}{\overline{b}}              
\newcommand{\og}{\overline{g}^a}            
\newcommand{\Rmu}{{\cal R}_\mu}             
\newcommand{\fa}{a}                         
\newcommand{\fb}{b}                         
\newcommand{\fan}{a_n}                      
\newcommand{\fbn}{b_n}                      
\begin{document}
\draft

\title{Ginsparg--Wilson Relation and Ultralocality\cite{grant}}
\author{Ivan Horv\'ath\cite{email}}
\address{ 
Department of Physics, University of Virginia\\
Charlottesville, Virginia 22903, USA}
\date{\today}
\maketitle


\begin{abstract}
It is shown that it is impossible to construct a free theory of fermions
on infinite hypercubic Euclidean lattice in four dimensions that is:
(a) ultralocal, (b) respects symmetries of hypercubic lattice,
(c) corresponding kernel satisfies 
${\bf D}\gfive + \gfive{\bf D} = {\bf D}\gfive{\bf D}$
(Ginsparg-Wilson relation),   
(d) describes single species of massless Dirac fermions in the continuum
limit.
\end{abstract}

The quest for incorporating chiral symmetry in lattice regularized gauge
theory has a long history. Many approaches with various degrees of depth and
{\ae}sthetic appeal have been tried. Even though a remarkable progress
has been achieved during the last few years, it is hard to match a striking
elegance and clarity of the picture that emerged during the last few months.
So far, these developments are mostly relevant for the vectorlike case 
but applications to chiral gauge theories appear to be around the corner.

At the heart of the above progress was the
realization~\cite{Has98A,Has98B,Neu98A} that there exist potentially
viable actions, satisfying Ginsparg--Wilson (GW) relation~\cite{Gin82A}.
If ${\bf D}$ is a lattice Dirac kernel, then the simplest form of
GW relation is
\begin{equation}
     \gfive {\bf D} + {\bf D} \gfive = {\bf D} \gfive {\bf D} \,.
     \label{eq:in10}
\end{equation}
L\"uscher recognized that (\ref{eq:in10}) can be viewed as a symmetry 
condition \cite{Lus98A}. Unlike chiral symmetry, this new symmetry
(which we call Ginsparg-Wilson-L\"uscher (GWL) symmetry)
involves a transformation that couples variables on different
lattice sites and becomes a standard chiral symmetry only in the continuum
limit. GWL symmetry has virtually the same consequences for the dynamics of
the lattice theory as the chiral symmetry does on the dynamics in continuum.
In particular, it guarantees the correct anomaly structure and the current
algebra predictions for low energy QCD directly on the
lattice~\cite{Has98B,Gin82A,Lus98A,Cha98A}. 
There is no need for tuning to recover aspects of chiral symmetry,
there are no complicated renormalizations, and there is no mixing between
operators in different chiral representations~\cite{Has98B}. Particularly 
striking are also the completely new avenues for studying topology on the
lattice~\cite{Has98A,Cha98A,Tin98A}. All of this can be discussed in the 
standard local field theory framework as a consequence of GWL symmetry.
In view of the Nielsen-Ninomiya theorem~\cite{Nie81A}, it is hard to imagine
to have things any better than this with respect to chiral issues in lattice
QCD.

While all of this definitely holds a promise of an extraordinary progress
in the near future, the troubling history of chiral symmetry on the lattice 
indicates that it might come for a price. Indeed, one drawback of the known
solutions of the GW relation is that they are not ultralocal, i.e. that
the interaction between fermionic variables is nonzero for sites 
arbitrarily far away from each other~\cite{Bie98A}. This complicates
the perturbation theory with
such actions considerably and, also, one looses the obvious numerical 
advantages steming from sparcity of the conventional operators such as
Wilson-Dirac operator. Moreover, while locality (exponential decay of
interaction at large distances) can be ensured easily for free actions,
it is usually not obvious in the presence of the gauge fields if the action
is not ultralocal. Consequently, it would be much preferable to work with
ultralocal actions and the question arises whether GWL symmetry and
ultralocality can at least in principle coexist.

In this Letter, it is argued that such hopes may not materialize.
In particular, we prove that the GW relation (\ref{eq:in10}) can not
be satisfied by a free ultralocal kernel defining a theory
with appropriate continuum limit, and respecting the symmetries of the
hypercubic lattice. Such a theorem can be extended to more general
GW relation  
$\gfive {\bf D} + {\bf D} \gfive = 2{\bf D} \gfive {\bf R}{\bf D}$,
with ${\bf R}$ being an ultralocal matrix, trivial in spinor space.
Also, analogous statements hold in two dimensions. Discussion of these
results as well as more detailed account of the proof presented here
will be given elsewhere~\cite{Hor98A}. 

Consider a system of $4$-component fermionic degrees of freedom living
on the sites of an infinite $4$-dimensional hypercubic Euclidean lattice.
Free theory of these fermions is described by some kernel ${\bf D}$ which
can be uniquely expanded in the form
\begin{equation}
    {\bf D}_{m,n} = \sum_{a=1}^{16} {\bf G}^a_{m,n}\Gamma^a.
    \label{eq:fi10}
\end{equation}
In the above equation $m,n$ label the space-time lattice points and 
$\Gamma^a$-s are the elements of the Clifford basis 
$\Gamma\equiv\{\identity,\gmu,\gfive,\gfive\gmu,\sigma_{\mu\nu,(\mu<\nu)}\}$.
Gamma--matrices satisfy anticommutation relations
$ \{ \gmu,\gnu \} = 2\dmunu\identity$, and we define 
$\gfive=\gone\gtwo\gamma_3\gamma_4$,
$\sigma_{\mu\nu}\equiv {i\over 2}[\gmu,\gnu]$.
Because of the completeness of Clifford basis on the space of
$4\times 4$ complex matrices, Eq.~(\ref{eq:fi10}) describes arbitrary
kernel and thus arbitrary quadratic action $\bar\psi{\bf D}\psi$.

The requirements of symmetry, ultralocality and continuum limit constitute
the set of restrictions on the above space of fermionic lattice theories.
Before we proceed to implement them, let us note that hypercubic lattice
structure is invariant under translations by arbitrary lattice vector
and under the subset of $O(4)$ transformations -- hypercubic rotations
and reflections. We refer to the former as {\it translation invariance}
and to the latter as {\it hypercubic invariance}.

{\em Translation Invariance and Ultralocality:} 
Translation invariance restricts the form of the action substantially
since it implies
\begin{displaymath}
    {\bf G}_{m,n}^a = {\bf G}_{0,n-m}^a\equiv g_{n-m}^a \qquad\quad
    a=1,2,\ldots 16.
\end{displaymath}
By ultralocality we mean that the fermionic variables do not interact 
beyond some finite lattice distance. Let us denote by ${\cal C}_N$ the
set of all lattice sites contained in the hypercube of side $2N$, centered 
at $n=0$, i.e. ${\cal C}_N\equiv\{n : |n_\mu|\le N,\, \mu=1,\ldots,4\}$.
One convenient way of defining ultralocality for translationally invariant
actions is to require the existence of a positive integer $N$, so that
$g_n^a = 0 \,, \;\forall n \not\in {\cal C}_N \;, \forall a$.
Translation invariance and ultralocality together imply the existence of
the diagonal Fourier image of the space--time part of ${\bf D}$.
In particular,
\begin{equation}
    D(p) = \sum_{a=1}^{16} G^a(p)\Gamma^a\,,
    \label{eq:fi20}
\end{equation}
where
\begin{equation}
    G^a(p)\equiv \sum_{n\in{\cal C}_N} g_n^a e^{ip\cdot n} \,.
    \label{eq:fi30}
\end{equation}
$G^a(p)$ are thus the complex--valued periodic functions of lattice
momenta $p\equiv (p_1,\ldots,p_4)$, whose Fourier series has a {\em finite}
number of terms. It should be emphasized that there is a true mathematical
equivalence between the set of all kernels satisfying translation invariance
and ultralocality, and the set of kernels 
defined by Eqs.~(\ref{eq:fi20},\ref{eq:fi30}).

{\em Hypercubic Symmetry:} We will discuss hypercubic symmetry directly in
the Fourier space which is convenient for our purposes.
Let ${\cal H}$ be an element of the hypercubic group in defining
representation and $H$ the corresponding element of the representation
induced on hypercubic group by spinorial representation of $O(4)$. 
We require that the action $\bar\psi D \psi$ does not change under
$\psi(p)\longrightarrow H\psi({\cal H}^{-1}p)$, 
$\bar\psi(p)\longrightarrow \bar\psi({\cal H}^{-1}p)H^{-1}$. 
This is equivalent to the requirement 
\begin{equation}
   D(p) = \sum_{a=1}^{16} G^a(p)\Gamma^a = 
          \sum_{a=1}^{16} G^a({\cal H}p)H^{-1}\Gamma^aH.
   \label{eq:fi35}
\end{equation}    
Since any hypercubic transformation ${\cal H}$ can be decomposed into 
products of reflections of single axis (${\cal R}_\mu$) and exchanges
of two different axis (${\cal X}_{\mu\nu}$), it is sufficient to require
invariance under these operations. Transformation properties of all the
elements of the Clifford basis are determined by the fact that $\gmu$
transforms as $p_\mu$ (vector). In particular
\begin{displaymath}
  R_\nu^{-1}\gmu R_\nu \;=\; \cases{-\gmu,& if $\mu=\nu$;\cr
                                     \gmu,& if $\mu\ne\nu$,\cr}
\end{displaymath}
and
\begin{displaymath}
  X_{\rho\sigma}^{-1}\gmu X_{\rho\sigma} \;=\;
                 \cases{\gamma_\sigma,& if $\mu=\rho$;\cr
                        \gamma_\rho,&   if $\mu=\sigma$;\cr
                        \gmu,&          otherwise.\cr}
\end{displaymath}
The elements of the Clifford basis naturally split into groups with
definite transformation properties and the hypercubic symmetry thus
translates into definite algebraic requirements on functions $G^a(p)$
some of which we will exploit.

{\em Continuum Limit:} Next, it is required that low energy physics happens
only for $p\sim 0$, where it corresponds to a single massless relativistic
Dirac fermion in the continuum. This implies the following local properties 
\begin{equation}
   G^a(p) = \cases{ip_\mu + O(p^2),&if $\Gamma^a  =\gmu\,$;\cr
                            O(p^2),&if $\Gamma^a\ne\gmu\,,\forall\mu\,$,\cr}
   \label{eq:fi40}
\end{equation}
and the restriction that $D(p)$ has to be invertible away from the origin
of the Brillouin zone (no doublers).

We now put forward the following definition:
\medskip

{\bf Definition (Set ${\cal U}$)}  
{\it Let $\mu\in\{1,2,3,4\}$, and let $a\in\{1,2,\ldots,16\}$. Let further
     $G^a(p)$ are the complex valued functions of real variables $p_\mu$,
     and let $D(p)$ be the corresponding matrix function constructed as in
     Eq.~{\rm (\ref{eq:fi20})}. 
     We say that the $16$-tuple $(G^1,\ldots,G^{16})$ belongs to the set 
     ${\cal U}$ if and only if the following holds:
     \begin{description}
       \item[$(\alpha)$] $\exists\; {\cal C}_N$ such that $G^a(p)$ has
             the form {\rm (\ref{eq:fi30})}, $\forall a$.
       \item[$(\beta)$] $D(p)$ satisfies condition {\rm (\ref{eq:fi35})}.
       \item[$(\gamma)$] $D(p)$ satisfies: $D\gfive + \gfive D = D\gfive D$.
       \item[$(\delta)$] $G^a(p)$ satisfy {\rm (\ref{eq:fi40})}, $\forall a$.
       \item[$(\epsilon)$] $D(p)$ is invertible unless
              $p_\mu=0\pmod{2\pi},\, \forall\mu$.  
     \end{description}
}
\medskip
\noindent It should be noted that for every action (kernel) (\ref{eq:fi10})
satisfying our requirements there is a corresponding element of ${\cal U}$ and
vice-versa. If the requirement of ultralocality is replaced by a weaker
condition of locality (at least exponential decay at large distances), then 
there exist free actions satisfying the rest of the conditions. However,
there do not appear to be examples of ultralocal actions enjoying the same
level of symmetry. In fact, we now prove the following statement:

\bigskip
{\bf Theorem} {\it Set ${\cal U}$ is empty.}
\medskip

\noindent{\it Proof.}
We will proceed by contradiction. Assume that there is at least one element
$(G^1,\ldots,G^{16}) \in {\cal U}$. To such an element we can assign
a $16$-tuple of functions of {\it single variable} 
$(\OG^1,\ldots,\OG^{16})$ by restricting $G^a(p)$ to the points
$p\equiv (q,q,0,0)$, i.e.
\begin{displaymath}
  G^a(p)\;\; 
  {\smash{\mathop{\hbox to 55pt{\rightarrowfill}}\limits^{p=(q,q,0,0)}}}
      \;\; \OG^a(q).
\end{displaymath}
We now investigate the consequences of conditions $(\alpha)-(\epsilon)$
on restrictions $\OG^a(q)$.

$(\alpha)$ As a consequence of Eq.~(\ref{eq:fi30}), functions $\OG^a(q)$
have Fourier series with finite number of terms i.e. there exist
non--negative integers $K,L$, such that
\begin{equation}
    \OG^a(q) = \sum_{-L\le m \le K} \og_m e^{iq\cdot m }, \;\;\forall a\,.
    \label{eq:fi50}
\end{equation} 

$(\beta)$ Consider the terms in $D(p)$ of the form $iB_\mu(p)\gmu$.
Invariance under reflections implies
\begin{displaymath}
  B_\mu(\ldots,-p_\nu,\ldots)= 
      \cases{-B_\mu(\ldots,p_\nu,\ldots),\,& $\mu=  \nu$;\cr
             +B_\mu(\ldots,p_\nu,\ldots),\,& $\mu\ne\nu$.\cr}
\end{displaymath}
Applying this to the reflection of $p_3$ or $p_4$ we have
\begin{displaymath} 
  \OB_4(q)=B_4(q,q,0,0)=-B_4(q,q,0,0)=0\,,
\end{displaymath}
and similarly $\OB_3(q)=0$. Furthermore, since under ${\cal X}_{12}$,
$\gone$ exchanges with $\gtwo$ we must have 
\begin{displaymath}
   B_1(q,q,0,0)=B_2(q,q,0,0)\equiv \OB(q).
\end{displaymath}
Next, consider the term $C(p)\gfive$. 
Since $\gfive\rightarrow -\gfive$ under $\Rmu$, it is required that
\begin{displaymath}
   C(\ldots,-p_\mu,\ldots)=-C(\ldots,p_\mu,\ldots),\;\forall\mu\,.
\end{displaymath}
Reflecting the component $p_4$, for example, this gives
\begin{displaymath}
  \overline{C}(q)=C(q,q,0,0)=-C(q,q,0,0)=0.
\end{displaymath}  
For the terms of the form $iE_\mu(p)\gfive\gmu$ invariance under
reflections demands
\begin{displaymath}
  E_\mu(\ldots,-p_\nu,\ldots)= 
      \cases{+E_\mu(\ldots,p_\nu,\ldots),& $\mu=  \nu$;\cr
             -E_\mu(\ldots,p_\nu,\ldots),& $\mu\ne\nu$,\cr}
\end{displaymath}
and using similar arguments as above, we can infer from this that
$\overline{E}_\mu(q)=0,\,\forall\mu$.
Finally, consider the terms $F_{\mu\nu}\sigma_{\mu\nu}$. Invariance under
reflections implies
\begin{displaymath}
  F_{\mu\nu}(\ldots,-p_\rho,\ldots)= 
      \cases{-F_{\mu\nu}(\ldots,p_\rho,\ldots),&$\rho= \mu$ or $\nu$;\cr
             +F_{\mu\nu}(\ldots,p_\rho,\ldots),&otherwise,\cr}
\end{displaymath}
which in turn ensures that $\overline{F}_{\mu\nu}(q)=0$, except for
$\overline{F}_{12}(q)$. However, under the exchange ${\cal X}_{12}$ of
$p_1$ and $p_2$ we have $\sigma_{12}\rightarrow -\sigma_{12}$, while  
$\overline{F}_{12}(q) \rightarrow \overline{F}_{12}(q)$, implying that
even this term has to vanish.
Summarizing the relevant implications of hypercubic symmetry, restriction
$\overline{D}(q)$ of $D(p)$ must have the form
\begin{equation}
    \overline{D}(q) = \Bigl( 1- \OA(q) \Bigr) \identity 
                      + i\OB(q)\Bigl(\gone + \gtwo \Bigr) \,.
    \label{eq:fi60}
\end{equation}

$(\gamma)$ GW relation for $\overline{D}(q)$ given in Eq.~(\ref{eq:fi60})
takes a simple form
\begin{equation}
   \OA^2 + 2\,\OB^2 = 1.
   \label{eq:fi70}
\end{equation}

$(\delta)$ The local properties~(\ref{eq:fi40}) imply 
\begin{equation}
   \OA(q) = 1 + O(q^2) \qquad \OB(q) = q + O(q^2).
     \label{eq:fi80}
\end{equation}

To proceed, we will rely on the Lemma stated bellow this proof. According 
to the Lemma, the solutions of equation (\ref{eq:fi70}) that have the form
(\ref{eq:fi50}) (with some minimal $K,L$) only exist if $K=L$. 
If $K=L=0$ (case of constant functions),
then condition (\ref{eq:fi80}) can not be satisfied and to avoid 
contradiction, we have to assume that $K=L>0$. If that is the case, 
then the Lemma states that the necessary (but not sufficient) condition for
$(\OA,\OB)$ to be the solution of equation (\ref{eq:fi70}) is that only the 
highest frequency modes are present in their Fourier expansion, i.e.
\begin{displaymath}
    \OA(q) = \oa_{-K} e^{-iq\cdot K} + \oa_K e^{iq\cdot K}\,,
\end{displaymath}
\begin{displaymath}
    \OB(q) = \ob_{-K} e^{-iq\cdot K} + \ob_K e^{iq\cdot K}\,.
\end{displaymath}
Conditions (\ref{eq:fi80}) then dictate uniquely what the coefficients in
the above equations have to be. In particular, $\oa_{-K}=\oa_K=1/2$
and $\ob_{-K}=-\ob_K=i/2K$, which corresponds to
$\OA(q)=\cos(Kq),\,\OB(q)=\sin(Kq)/K$. For these functions we have
\begin{displaymath}
 \OA^2 + 2\,\OB^2 \;=\; \cos^2(Kq) + {2\over {K^2}}\sin^2(Kq)\,,
\end{displaymath}
and consequently, equation (\ref{eq:fi70}) can only be satisfied if
$2/K^2=1$. However, there is no positive integer $K$ so that this
condition is satisfied. We have therefore arrived at the contradiction with
the existence of $(G^1,\ldots,G^{16}) \in {\cal U}$ and the proof is thus
complete.

$\Box$

In essence, the above proof relies on two major ingredients: First is the
fact that it is sufficient to consider a single periodic direction in the
Brillouin zone and that the hypercubic symmetry is powerful enough to render
the problem tractable. The second ingredient is a perhaps surprising result
that periodic solutions of equations of type (\ref{eq:fi70}) either
involve a single Fourier component or infinitely many of them. 
This is summarized by the following Lemma, whose complete proof will
be given in the detailed account of this work~\cite{Hor98A}:
\medskip

 {\bf Lemma}
 {\it Let $K$,$L$ be nonnegative integers and $d$ a positive real number.
  Consider the set ${\cal F}^{K,L}$ of all pairs of functions 
  $(\,\A(q),\B(q)\,)$ that can be written in form
  \begin{displaymath}
     \A(q) = \sum_{-L\le n \le K} \fan  e^{iq\cdot n} \qquad
     \B(q) = \sum_{-L\le n \le K} \fbn  e^{iq\cdot n} \;,
  \end{displaymath}
  where $q\in\R ,\, n\in\Z$, and $\fan,\fbn\in\C$ are such that $\fa_K,\fb_K$
  do not vanish simultaneously and $\fa_{-L},\fb_{-L}$ do not vanish 
  simultaneously. Further, let ${\cal F}^{K,L}_d \subset {\cal F}^{K,L}$
  denotes the set of all solutions on ${\cal F}^{K,L}$ of the equation
  \begin{equation}
     \A(q)^2 + d\,\B(q)^2 = 1\;.
     \label{eq:app20}
  \end{equation} 
  Then the following holds:
  \smallskip

  \noindent {\rm (a)} If $K=L=0$, then
  \begin{displaymath} 
     {\cal F}^{0,0}_d = 
        \{\,(\fa_0,\fb_0)\,:\, \fa_0^2 + d\,\fb_0^2=1\, \}\,.
     \end{displaymath}

  \noindent {\rm (b)} If $K=L>0$, then 
  $\;{\cal F}^{K,K}_d = \{\,(\,\A(q),\B(q)\,)\,\}$,
  with
  \begin{eqnarray*}
     \A(q) &=& \fa_{-K}\,e^{-iq\cdot K} + \fa_K\,e^{iq\cdot K} \\
     \B(q) &=& \fb_{-K}\,e^{-iq\cdot K} + \fb_K\,e^{iq\cdot K}\,, 
  \end{eqnarray*}
  and  
  \begin{displaymath}
     \fa_K = \,c\,i\sqrt{d}\,\fb_K \qquad
     \fa_{-K}  =  {c\over {4i\sqrt{d}\,\fb_K}}\qquad
     \fb_{-K}  = {1\over {4d\,\fb_K}}\,,
  \end{displaymath} 
  where $\fb_K\ne 0, \;\sqrt{d}>0$ and $c=\pm 1$.
  \medskip

  \noindent{\rm (c)} If $K\ne L$, then $\;{\cal F}^{K,L}_d = \emptyset$.
} 
\medskip

\noindent {\it Outline of the proof:}
Using the completeness and othogonality of the Fourier basis,
equation (\ref{eq:app20}) is equivalent to the following set of conditions
on Fourier coefficients  
\begin{displaymath}
     \sum_{\scriptstyle -L\le n\le K \atop
           \scriptstyle -L\le k-n\le K} \fa_n\fa_{k-n} \;+\;
     d\sum_{\scriptstyle -L\le n\le K \atop
            \scriptstyle -L\le k-n\le K} \fb_n\fb_{k-n} \;=\; \delta_{k,0} \;,
\end{displaymath} 
where $-2L\le k \le 2K$.

Case (a) is obvious and we start with case (b):
The idea is to explicitly solve the above equations by analyzing
them in appropriate sequence. We start with the group $\;K\le k \le 2K$,
which only involves coefficients of non--negative frequencies. By induction,
starting from $k=2K$ and continuing down, it is possible to show that
this group of conditions is equivalent to
\begin{equation}
    \fa_n = \,c\,i\sqrt{d}\,\fb_n \qquad
    \sqrt{d}>0,\; c=\pm 1\,,
    \label{eq:app60} 
\end{equation}   
where $n=0,1,\ldots,K$.
Similarly, analyzing the group involving only coefficients of non--positive
frequencies, i.e. $\;-2K\le k\le -K$, we arrive at
\begin{equation}
    \fa_{-n} = \,\cbar\,i\sqrt{d}\,\fb_{-n} \qquad
    \sqrt{d}>0,\; \cbar =\pm 1\,,
    \label{eq:app70} 
\end{equation}   
for $n=0,1,\ldots,K$. Inserting results (\ref{eq:app60},\ref{eq:app70})
in condition for $k=0$, implies $\cbar=-c$, and consequently,
\begin{displaymath}
    \fa_0 \,=\,\fb_0 \,=\,0.
\end{displaymath} 
Using these results, we can start induction at $k=K-1$ to show that 
conditions for $\;1\le k\le K-1$ lead to
\begin{displaymath}
   \fb_{-n} = 0 = \fa_{-n}    \qquad\quad n=1,2,\ldots ,K-1\,,
\end{displaymath}
and, analogously, for $\;-K+1\le k\le -1$ we arrive at
\begin{displaymath}
   \fb_n = 0 = \fa_n   \qquad\quad n=1,2,\ldots ,K-1 \,.
\end{displaymath}
Finally, the last condition that was not fully exploited is the one
for $k=0$, which now simplifies to
\begin{displaymath}
   \fb_K\fb_{-K} \,=\, {1\over{4d}}\,.
\end{displaymath}
The above steps establish the result (b).

Case (c): Technically, this is arrived at in a completely analogous manner 
to case (b). However, due to the asymmetry between positive and negative 
frequencies, the equation for $k=0$ can never be satisfied.  

$\Box$

Let us close by noting that in the proof of the Theorem, condition
$(\epsilon)$ was not used at all. In other words, there are no acceptable 
ultralocal solutions of (\ref{eq:in10}) with or without doublers. This is
not true if the requirement of hypercubic symmetry is relaxed. 
In that case, there exist ultralocal solutions with doublers 
and it is still an open question whether doubler--free solutions
do exist. Since breaking the hypercubic symmetry carries
with itself the necessity of tuning to recover rotation
invariance in the continuum limit, it is not obvious whether such a 
possibility would be practically viable. On the other hand, theoretically
it would be quite interesting to know whether hypercubic symmetry can
be traded for GWL symmetry.

The author thanks Wolfgang Bietenholz, Mike Creutz, Peter Hasenfratz,
Robert Mendris, Martin Moj\v{z}i\v{s}, Tony Kennedy and Hank Thacker, whose 
input was useful at various stages of this work.


\begin{references}

\bibitem[*]{grant} This work was supported in part by the U.S. Department of
Energy under grant DE-AS05-89ER40518.

\bibitem[\dagger]{email} e--mail: ih3p@virginia.edu

\bibitem{Has98A}
P.~Hasenfratz, V.~Laliena and F.~Niedermayer,\nobreak
Phys.~Lett. {\bf B427} (1998) 125.

\bibitem{Has98B}
P.~Hasenfratz, hep-lat/9802007.

\bibitem{Neu98A}
H.~Neuberger, Phys.~Lett. {\bf B427} (1998) 353.

\bibitem{Gin82A}
P.~Ginsparg and K.~Wilson, Phys.~Rev. {\bf D25} (1982) 2649.

\bibitem{Lus98A}
M.~L\"{u}scher, Phys.~Lett. {\bf B428} (1998) 342.

\bibitem{Cha98A}
S.~Chandrasekharan, hep-lat/9805015.

\bibitem{Tin98A}
T.~W.~Chiu and S.~V.~Zenkin, hep-lat/9806019;\\
T.~W. Chiu, hep-lat/9804016.

\bibitem{Nie81A}
H.~B.~Nielsen and M.~Ninomiya, Phys.~Lett. {\bf B105} (1981) 219;
Nucl.~Phys. {\bf B185} (1981) 20[E: B195 (1982) 541].

\bibitem{Bie98A}
W.~Bietenholz, hep-lat/9803023.

\bibitem{Hor98A}
I.~Horv\'ath, in preparation.

\end{references}
\end{document}